\documentclass[journal]{IEEEtran}
\usepackage{ifpdf}
\usepackage{bm}
\usepackage{cite}
\ifCLASSINFOpdf
\usepackage{graphicx}
\else
\fi
\usepackage{amsmath}
\usepackage{amsfonts}
\usepackage{array}
\usepackage{url}
\usepackage{booktabs}
\usepackage{arydshln}
\usepackage{algorithm}
\usepackage[noend]{algpseudocode}
\usepackage{amsfonts}
\usepackage{tabularx}
\usepackage{amssymb}
\usepackage{multirow}
\usepackage{threeparttable}
\usepackage{makecell}
\usepackage{rotating}
\usepackage{multicol}
\usepackage{booktabs}
\usepackage{xcolor}
\usepackage{array}
\usepackage{subfigure}
\usepackage{verbatim}
\usepackage{lineno,hyperref}
\usepackage{amsmath,bm}
\newcommand\our{\textsc{ProgRE}}

\begin{document}

\title{Progressive Residual Extraction based Pre-training for Speech Representation Learning}

\author{Tianrui Wang, Jin Li, Ziyang Ma, Rui Cao, Xie Chen, Longbiao Wang,  Meng Ge\\Xiaobao Wang,  Yuguang Wang, Jianwu Dang, Nyima Tashi
\thanks{
This work was supported in part by the National Key R\&D Program of China (2022ZD0116101), the National Natural Science Foundation of China under Grant U23B2053 and Grant 62176182. \textit{(Corresponding authors: Longbiao Wang and Meng Ge)}.

Tianrui Wang, Jin Li, Rui Cao, and Xiaobao Wang are with the Tianjin Key Laboratory of Cognitive Computing and Application, College of Intelligence and Computing, Tianjin University, Tianjin 300350, China (e-mail: \{wangtianrui, caorui\_2022, lijin0120, wangxiaobao\}@tju.edu.cn).

Longbiao Wang is with the Tianjin Key Laboratory of Cognitive Computing and Application, College of Intelligence and Computing, Tianjin University, Tianjin 300350, China, and also with the Huiyan Technology (Tianjin) Company, Ltd., Tianjin 300350, China (e-mail: longbiao\_wang@tju.edu.cn).

Meng, Ge is with Saw Swee Hock School of Public Health, National University of Singapore, Singapore (e-mail: gemeng@nus.edu.sg).

Ziyang Ma and Xie Chen are with MoE Key Lab of Artificial Intelligence, AI Institute, Shanghai Jiao Tong University, Shanghai, China (e-mail: \{zym.22, chenxie95\}@sjtu.edu.cn).

Yuguang Wang is with Huiyan Technology (Tianjin) Co., Ltd, Tianjin, China (e-mail: ygwang@huiyan-tech.com).

Jianwu Dang is with the Shenzhen Institute of Advanced Technology, Chinese Academy of Sciences, Guangdong, China (e-mail: jdang@jaist.ac.jp).

Nyima Tashi is with the School of Information Science and Technology, Tibet University, Lhasa, China (e-mail: nmzx@utibet.edu.cn)

}}

\markboth{IEEE Transaction of Audio, Speech and Language Processing}%
{Shell \MakeLowercase{\textit{et al.}}: Multimodal word discovery with spoken descriptions and visual concepts}

\maketitle

\begin{abstract}


Self-supervised learning (SSL) has garnered significant attention in speech processing, excelling in linguistic tasks such as speech recognition. 
However, jointly improving the performance of pre-trained models on various downstream tasks, each requiring different speech information, poses significant challenges. To this purpose, we propose a progressive residual extraction based self-supervised learning method, named \our{}. Specifically, we introduce two lightweight and specialized task modules into an encoder-style SSL backbone to enhance its ability to extract pitch variation and speaker information from speech. Furthermore, to prevent the interference of reinforced pitch variation and speaker information with irrelevant content information learning, we residually remove the information extracted by these two modules from the main branch. The main branch is then trained using HuBERT's speech masking prediction to ensure the performance of the Transformer's deep-layer features on content tasks. In this way, we can progressively extract pitch variation, speaker, and content representations from the input speech. Finally, we can combine multiple representations with diverse speech information using different layer weights to obtain task-specific representations for various downstream tasks. Experimental results indicate that our proposed method achieves joint performance improvements on various tasks, such as speaker identification, speech recognition, emotion recognition, speech enhancement, and voice conversion, compared to excellent SSL methods such as wav2vec2.0, HuBERT, and WavLM.
    
\end{abstract}

\begin{IEEEkeywords}
    Self-supervised Learning, Speech Representation Learning, Speech Disentangle, Pre-training
\end{IEEEkeywords}

\IEEEpeerreviewmaketitle

\section{Introduction}
\label{intro}
Speech self-supervised learning (SSL) aims to learn how to extract a universal representation of speech for various downstream tasks based on a massive amount of unlabeled data \cite{apc}. In this framework, a model is pre-trained on tasks using the speech itself to generate supervisory signals, rather than relying on external labels provided by humans \cite{ssl_review}. After pre-training, the model, regarded as a speech representation extractor, is fine-tuned using supervised speech data to achieve task-specific capabilities for specific downstream tasks \cite{superb}.

Existing well-known speech SSL methods can be categorized into two streams: generative and contrastive methods \cite{ssl_cate}. 
Generative methods build an encoder to convert speech into representations and train the encoder by reconstructing the speech from these representations, including TERA \cite{tera}, SoundStream \cite{soundstream}, and Encodec \cite{encodec}. Since generative methods are supervised on specific speech signals, they often excel in acoustic tasks but are not satisfied for content tasks \cite{viola}.
Contrastive methods also build an encoder to convert speech into representations, but train the encoder by measuring the similarity between representations of different inputs or modules. Examples include wav2vec \cite{wav2vec2}, HuBERT \cite{hubert}, WavLM \cite{wavlm}, and Data2vec \cite{data2vec2}. These contrastive methods are usually supervised on cluster-style macro information, so they perform well on content tasks but mediocrely on acoustic tasks \cite{usm,sslonse}.


With the development of multi-modal large language models \cite{zhang2023speechgpt}, the universality of SSL across various tasks has been highlighted \cite{universal_ssl}.
In pursuit of this objective, the SUPERB and SUPERB-SG benchmarks \cite{superb, superbsg} assemble fifteen downstream tasks to evaluate pre-trained models in areas, such as content, speaker, paralanguage, and acoustic processing. 
Although researchers have proposed various impressive SSL strategies tailored to specific tasks such as speaker recognition \cite{unispeechsat, dino}, emotion recognition \cite{emo2vec, emo2}, and speech enhancement \cite{wang2023adapter, se_ssl, lin2024selective}, enhancing a model's ability on one task often leads to a decline in its ability on other tasks \cite{ma2023pushing}. 
This prompts a challenging research question: Can speech pre-training be equipped with the capability to simultaneously enhance performance across various tasks?

To answer this research question, initial studies have focused on combining multiple task-specific pre-trained models \cite{multi_ssl} or using adapter-based multi-task pre-training \cite{wang2023adapter}. These researches facilitate the concurrent extraction of speech representations tailored for diverse downstream tasks, and achieve preliminary success. However, these approaches, rooted in the Mixture of Experts (MOE) principle \cite{moe}, lead to increased resource demands and do not fundamentally address the challenges inherent in achieving universal speech SSL representations. Some explorations have pointed out that the incompatibility between tasks makes it difficult for models to find a common direction of convergence across various tasks in multi-task learning \cite{wang2023adapter, lin2024selective}. 
The incompatibility between different tasks primarily stems from the varying contributions of different types of speech information to each task \cite{speechtokenizer}. Correspondingly, existing SSL models also exhibit a trend of task modularization when extracting speech representations. The representations produced by different layers are suited to different downstream tasks. Representations from the shallow layers of Transformers focus on capturing acoustic information, with these layers closer to the input modeling richer acoustic details \cite{layer_ana}. In contrast, deep-layer representations, which contain more contextual and semantic information, perform better on tasks such as speech recognition \cite{wav2vec_vc}. We refer to these as task characteristics in our paper.
The layer-wise task characteristics can be explained by speech information disentanglement. Speech can theoretically be progressively disentangled into non-linguistic, para-linguistic, and linguistic information \cite{fujisaki2004information}. 
In practice, speech is typically decoupled into three components: speaker, content, and pitch variation \cite{vqmivc, vc_review}. These three types of information are theoretically independent of each other and can be freely combined for use in various downstream tasks \cite{wav2vec_vc}.
Moreover, studies have indicated that removing content information can enhance speaker recognition performance \cite{spk_decouple, spk_decouple2, qian2022contentvec}. Conversely, other research \cite{speechtokenizer, giuliani2006improved} suggests that removing content-independent speaker information can improve the performance of content-related tasks.
Therefore, leveraging the independence between different types of speech information and the task-specific characteristics of various Transformer layers in SSL models seems to be the key to addressing the varying demands for speech information across different downstream tasks.

Inspired by the above discussions, we propose a novel pre-training method called \our{}, which progressively extracts the representations of pitch variation, speaker, and content from speech. 
By doing so, \our{} can ensure the extracted representations adapt to downstream tasks with various demands for speech information, achieving simultaneous compatibility effects.
Specifically, we first strengthen the extraction of pitch variation and speaker information in the two middle layers of the SSL model. 
Since pitch-variation, speaker, and content information are theoretically independent of each other \cite{vqmivc}, we progressively remove the strengthened pitch-variation and speaker representations from the main branch. 
This gradual purification of the main branch reduces the model's burden and prevents the strengthened information from interfering with the learning of other irrelevant information, especially content information.
Specifically, based on HuBERT \cite{hubert}, we insert two lightweight extractors to model pitch variation and speaker information of speech and progressively remove them from the main speech branch in a progressive residual manner \cite{encodec, soundstream, speechtokenizer}. 
Finally, the residual main branch is trained by HuBERT's self-supervised strategy to predict the masked units.
Experiments show that our \our{} can jointly improve performance across various tasks. 
Furthermore, visualizations demonstrate that specific layers contribute more significantly to their corresponding tasks. By strengthening the roles of different layers for different types of speech information, \our{} with a weighted-sum mechanism can also be used to analyze the downstream task's demands for various types of speech information.

Our main contributions are summarized as follows:
\begin{itemize}

\item We pointed out and experimentally verified that the task characteristics of different layers facilitated by the pre-training strategy, as well as mitigating the incompatibility between different tasks, are key to achieving a universal pre-training model.

\item We proposed a progressive residual extraction based pre-training method for speech representation learning. This approach enables the pre-trained model to balance the extraction of pitch variation, speaker information, and content information, leading to joint performance improvements across various downstream tasks.

\item We additionally introduced the extractors of pitch variation and speaker information, which can greatly improve the model's ability to extract intonation and non-linguistic information and achieve state-of-the-art (SOTA) performance on various downstream tasks, especially speaker identification and voice conversion tasks.

\item In addition to evaluating \our{}'s performance on a 960-hour dataset, we also validated its effectiveness on large-scale pre-training tasks using an 84,500-hour English-Chinese bilingual dataset. Furthermore, we released code at \url{https://github.com/wangtianrui/ProgRE}.

\end{itemize}

This paper is organized as follows. Section \ref{preliminary} introduces fundamental concepts utilized in our approach. Section \ref{pf} outlines the process by which our method extracts representations for various downstream tasks. Section \ref{method} provides a detailed description of our \our{} method. Section \ref{experiment_sec} presents the experimental setup and analysis of results. Section \ref{discussion} discusses the findings of the research. Finally, Section \ref{conclusion} concludes the paper.

\section{Preliminaries}
\label{preliminary}
\subsection{HuBERT}
\begin{figure}[htb]
\centering
\vspace{-0.45cm}
\includegraphics[width=9.0cm]{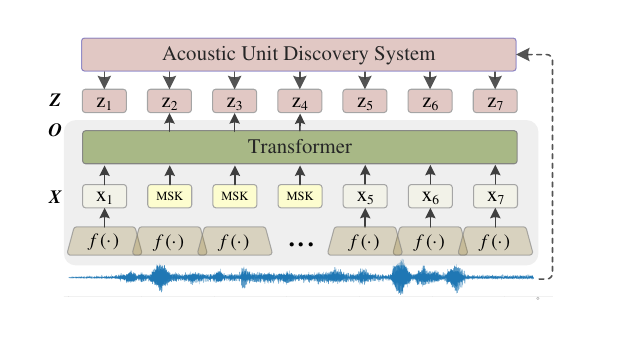}
\vspace{-1.0cm}
\caption{Diagram of HuBERT, which takes raw waveform as input to perform a BERT-like self-supervised pre-training.}
\vspace{-0.4cm}
\label{hubertpic}
\end{figure}

HuBERT \cite{hubert}, as shown in Fig.~\ref{hubertpic}, is a typical SSL method which benefits from an offline clustering to generate pseudo labels $\bm{Z}$ for a BERT-like pre-training \cite{bert}. A convolutional module $f(\cdot)$ converts signal into the frame-level feature $\bm{X}$. Then, the $\bm{X}$ is encoded by the Transformer into the representation $\bm{O}$. During pre-training, the frame-level features are masked randomly and successively, then fed to Transformer, the model is trained to predict the labels of the masked frames. 

\subsection{Residual Vector Quantization}
\label{rvq_sec}
\begin{figure}[htb]
\centering
\vspace{-0.65cm}
\includegraphics[width=9.cm]{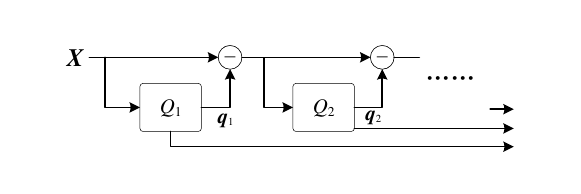}
\vspace{-1.0cm}
\caption{Diagram of residual vector quantization. RVQ performs progressive residual quantization of $\bm{X}$.}
\label{rvq_1}
\end{figure}

Residual vector quantization (RVQ) is commonly used in speech compression \cite{soundstream, encodec, speechtokenizer}, which performs progressively refined quantization of the representation $\bm{X}$, as shown in Fig~\ref{rvq_1}.

The representation $\bm{X}$ is first encoded by the first quantization $Q_1$, resulting in the representation $\bm{q}_1$. The error between $\bm{X}$ and $\bm{q}_1$ is then encoded by a second quantization module $Q_2$. In this way, the representation learned by $Q_2$ contains minimal information that was already captured by $Q_1$ \cite{speechtokenizer}.

\section{Problem Formulation}
\label{pf}
\begin{figure}[t]
\centering
\vspace{-0.6cm}
\includegraphics[width=8cm]{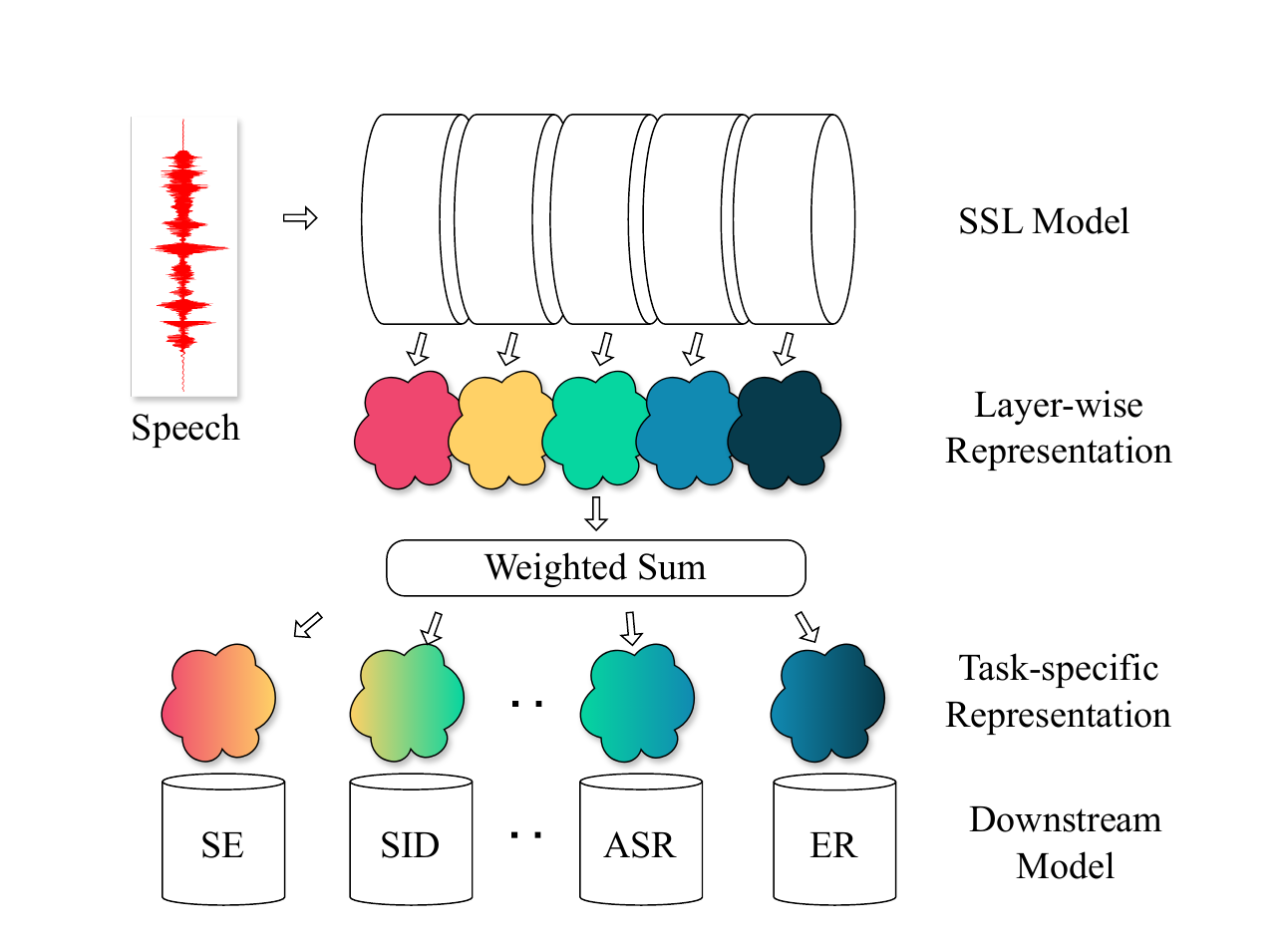}
\vspace{-0.5cm}
\caption{Diagram of the weighted-sum mechanism-based speech representation extraction. Speech is encoded into representations by a multi-layer SSL model, and then the task-specific representation for various downstream tasks is assembled with task-specific layer weights.}
\label{problem_formulation}
\end{figure}

Unlike conventional self-supervised pre-training models for speech, which aim to extract a single representation that can be widely used in various downstream tasks \cite{wav2vec, superb}, following SUPERB-SG \cite{superbsg}, our \our{} seeks to extract multiple representations of the input speech containing diverse speech information through a single SSL extractor. These representations can then be combined arbitrarily using a weighted-sum mechanism \cite{superbsg, wav2vec_vc} with minimal weights to obtain task-specific representations for various downstream tasks, as shown in Fig~\ref{problem_formulation}.

Given speech $\bm{x}$, \our{} uses a $n$-layer Transformer SSL model to extract layer-wise representations $\mathbf{O}=\left\{\bm{O}_1, \bm{O}_1, \dots, \bm{O}_n\right\}$. Then, the task-specific representation $\bm{R} \in \mathbb{R}^{T \times D}$ is reorganized as follows:
\begin{equation}
    \bm{R} = \sum_{i=1}^N \omega_i \cdot \bm{O}_i,
\end{equation}
where $\omega_i$ is the task-specific layer-wise weight. These weights can be learned from task-specific fine-tuning.

\begin{figure*}[t]
\centering
\vspace{-0.5cm}
\includegraphics[width=17cm]{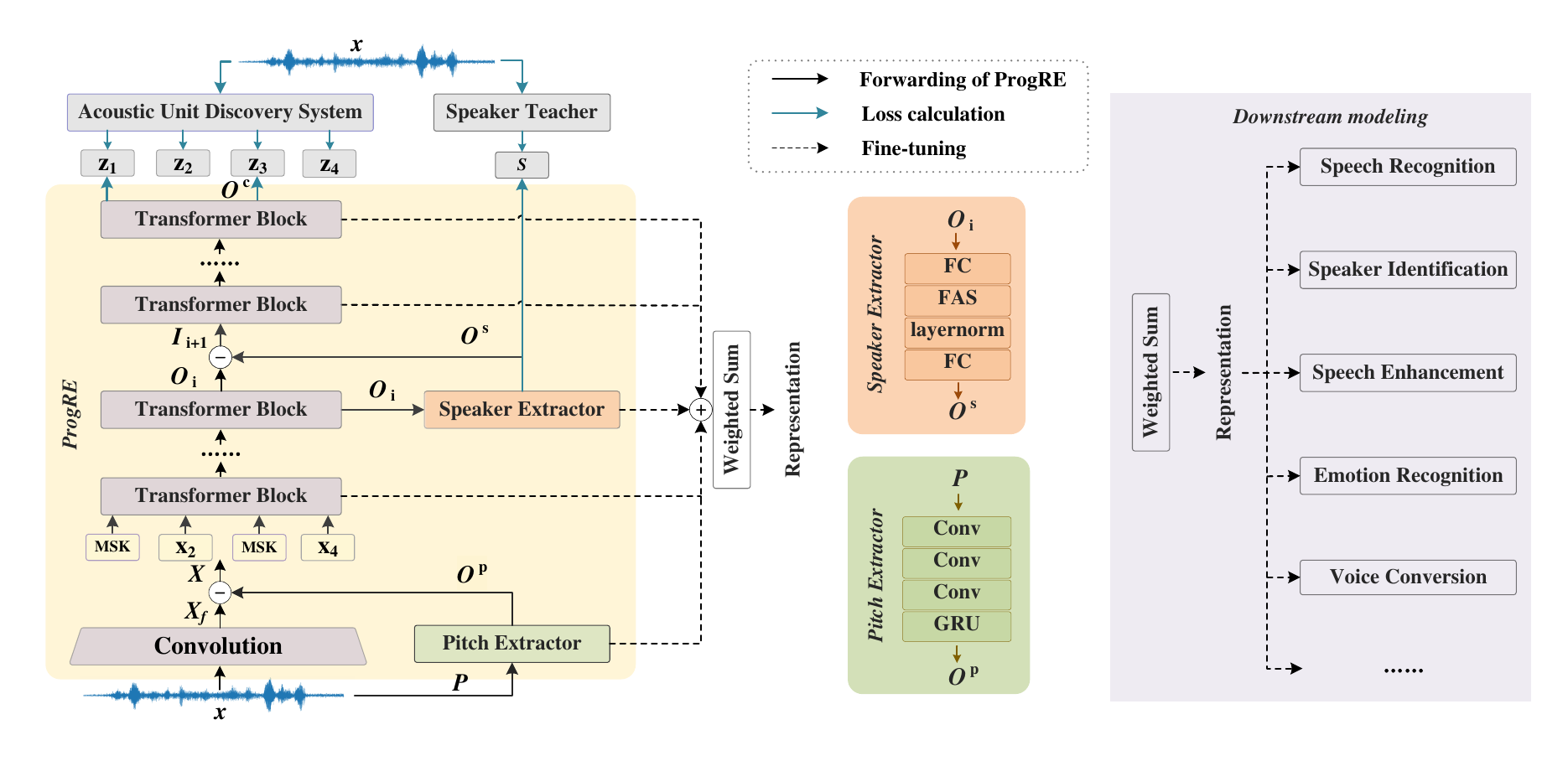}
\vspace{-0.8cm}
\caption{The diagram depicts our \our{} model, which takes a waveform as input and progressively extracts three types of representations: pitch variation $\bm{O}^\text{ p}$, speaker $\bm{O}^\text{ s}$, and content $\bm{O}^\text{ c}$ (indicated by black solid lines). The model is supervised by two offline systems trained on the unlabeled dataset (indicated by blue solid lines). For fine-tuning, a weighted-sum mechanism is employed (indicated by black dotted lines).} 
\vspace{-0.5cm}
\label{proposed}
\end{figure*}

\section{Progressive Residual Extraction Based Pre-training}
\label{method}
As mentioned in Section~\ref{intro}, the HuBERT pre-training strategy results in deeper Transformer layers extracting representations dominated by content information, while shallower layers retain more acoustic details. We propose a progressive residual extraction based scheme and adapt it into HuBERT, enhancing its ability to capture pitch variation and speaker information without compromising its outstanding performance in extracting content information.

As shown in Fig.~\ref{proposed}, 
the input waveform $\bm{x}$ is converted into a frame-level representation $\bm{X}_f$ with a frame stride of 20ms by a convolutional module consisting of 7 layers of 1-D convolution. Next, the pitch information $\bm{O}^{\text{ p}}$ is extracted by a pitch extractor and removed from the main branch to obtain $\bm{X}$. The multi-layer Transformer encodes $\bm{X}$ into the representation $\bm{O}$. In the middle $i$-th Transformer layer, we add a speaker extractor to extract the speaker information $\bm{O}^{\text{ s}}$ and remove it from the main branch. During pre-training, $\bm{X}$ is randomly masked before being input into the Transformer, and pseudo-labels for the main branch and the speaker teacher network are obtained based on unsupervised or self-supervised strategies. During fine-tuning, a weighted-sum mechanism is employed to obtain various features for downstream tasks, with learnable weights. We will introduce our \our{} method in detail in the following sub-sections.


\subsection{Progressive Residual Extraction}
\label{pre_sec}
In order to achieve information removal in our \our{}, we migrated Residual Vector Quantization (RVQ) mentioned in Section~\ref{rvq_sec} into continuous representation, performing residually refined extraction of the representation $\bm{X}$. We refer to this migrated method as progressive residual extraction.

As shown in the \our{} box in Fig.~\ref{proposed}, we adapted our progressive residual extraction method into the HuBERT framework. We inserted the pitch extractor and speaker extractor as continuous-version $Q_1$ and $Q_2$ of Fig.~\ref{rvq_1}, respectively, and removed the extracted representations from the main branch. The information in the main branch is progressively refined and finally supervised by HuBERT’s self-supervision strategy to learn the extraction of cluster information focusing on content \cite{layer_ana}. Since information removal in progressive residual extraction is only effective when all modules are trained jointly \cite{rvq_review}, all modules of \our{} are jointly pre-trained using HuBERT’s self-supervision strategy. Additionally, the speaker extractor is co-supervised by a teacher model specializing in capturing speaker information. Furthermore, the pitch extractor is constrained to extract only pitch variation information by inputting normalized pitch \cite{vqmivc}. In this way, \our{} can extract pitch variation representation, speaker representation, and content representation in a residual manner.

\subsubsection{Pitch Variation Modeling}
Following the speech decoupling approach of voice conversion \cite{vqmivc}, we extract the pitch $\bm{F}_0$ from the waveform as the anchor for intonation variation \cite{pitch_dio}. The representation of the pitch extractor is expected to contain intonation variations while excluding content and speaker information, so we then perform log-normalization \cite{normalization} within each waveform's pitch as follows:
\begin{equation}
    \bm{P} = \frac{\log{\bm{F}_0}-\text{mean}\left(\log{\bm{F}_0}\right)}{\text{std}\left(\log{\bm{F}_0}\right)}.
\end{equation}

We then use the normalized pitch as input to ensure that the representation extracted by the pitch extractor module only contains pitch variation information \cite{vqmivc}, thus ensuring the effectiveness of progressive residual extraction for other pitch-variation-irrelative tasks, such as speaker and content information extraction.
Specifically, we employ a lightweight convolutional recurrent module to process the normalized pitch, as illustrated in Fig.~\ref{proposed}. Each convolution block (Conv) consists of a 1D convolution, batch normalization \cite{normalization}, and ReLU activation function \cite{activations}. A single-layer GRU \cite{gru}, followed by an output fully connected (FC) layer, is utilized to extract the representation of pitch variation $\bm{O}^\text{ p}$.

Since the pitch is extracted from the waveform, we removed the pitch variation information from the main branch after the convolution block, as:
\begin{equation}
    \bm{X} = \text{layernorm}\left( \text{Convolution}\left( \bm{x} \right) - \bm{O}^\text{ p} \right),
\end{equation}
where $\bm{x}$ is the input signal, layer normalization is performed after removal to accelerate the convergence of the model \cite{normalization}.

\subsubsection{Speaker Information Modeling}
Unlike conventional utterance-level speaker representation extraction, the speaker extractor in \our{} is a frame-level extraction module. The frame-level module leverages the mask prediction pre-training strategy of SSL, enabling the encoder to learn to predict the information randomly masked in the input sequence, thereby improving the model’s bidirectional and global speaker information extraction ability.

The inserted speaker extractor comprises an FC layer, frame-level attentive statistics (FAS), and layer normalization following an output FC layer. FAS is a frame-level Attentive Statistic Pooling \cite{ecapa}, which calculates mean and variance on each frame. We insert the speaker extractor after the $i$-th Transformer block to extract the speaker representation $\bm{O}^{\text{ s}} \!=\! \left[\bm{o}^\text{ s}_1,\bm{o}^\text{ s}_2,\cdots,\bm{o}^\text{ s}_T\right] \in \mathbb{R}^{T\times D}$, as shown in Fig.~\ref{proposed}. 

In addition to being trained with the main branch using HuBERT's self-supervised strategy, we add a constraint to guide the speaker extractor in a teacher-student learning manner, focusing solely on speaker information. Similar to the K-means-based speech units of HuBERT, we obtain an utterance-level target $\bm{s} \in \mathbb{R}^K$ for the speaker extractor based on an offline self-supervised pre-trained model, EMA-DINO \cite{dino}, which is an ECAPA-TDNN \cite{ecapa} pre-trained without any labels using knowledge distillation. Then, masked regression is employed to train the speaker extractor as follows:
\begin{equation}
\mathcal{L}_\text{ s} = - \frac{1}{T_m} \sum_{t\in T_{m}} \log \sigma \left(\text{sim}( \bm{A}^s\bm{o}^\text{ s}_t, \bm{s} )\right),
\label{spk_loss}
\end{equation}
where $\bm{A}^s \in \mathbb{R}^{D \times K}$ is a projection matrix, $\text{sim}\left(\cdot,\cdot\right)$ represents the cosine similarity, and $\sigma(\cdot)$ denotes the sigmoid \cite{activations}. The speaker loss $\mathcal{L}_\text{ s}$ is calculated only on the masked frames $T_m$.

The extracted speaker representation $\bm{O}^{\text{ s}}$ is then removed from the output of the $i$-th Transformer block $\bm{O}_i$, as 
\begin{equation}
    \bm{I}_{i+1} = \text{layernorm}\left( \bm{O}_i - \bm{O}^{\text{ s}} \right),
\end{equation}
where $\bm{I}_{i+1}$ denotes the input of the next Transformer block.

\subsubsection{Content Information Modeling}
The training of the main branch in our model follows the HuBERT \cite{hubert}. Specifically, we employ the BERT-like masked pseudo-label prediction task \cite{bert} based on K-means clustering. This objective encourages the deep layers of the encoder to learn content representations while allowing the shallow-layer representations closer to the input to retain more acoustic details \cite{layer_ana}. Preserving these task characteristics of different layers is crucial for ensuring the effectiveness of the pitch and speaker extractors.

Before pre-training, we perform offline clustering of MFCC or hidden-layer representations from the previously pre-trained model to generate pseudo-labels $\bm{Z}\!=\!\left[z_1,z_2,\cdots,z_T\right]$, where each $z\in \left[U\right]$ is a $U$-class variable. As illustrated in Fig.~\ref{proposed}, during pre-training, the frame-level output of the convolution module is randomly and successively masked, and then fed into the Transformer encoder. After extracting and removing pitch variation and speaker information, the main branch is trained to predict the pseudo-labels of the masked frames, as:
\begin{equation}
    \mathcal{L}_\text{ c} = \frac{1}{T_m} \sum_{t\in T_{m}} \log {\frac{\exp\left(\text{sim}\left(\bm{A}^c\bm{o}^\text{ c}_t,\bm{e}_u\right)/\tau\right)}{\sum_{u'}^{U}\exp\left(\text{sim}\left(\bm{A}^c\bm{o}^\text{ c}_t,\bm{e}_{u'}\right)/\tau\right)}},
\label{hubert_loss}
\end{equation}
where $\bm{A}^c$ is a projection matrix, $\bm{O}^\text{ c}\!=\!\left[\bm{o}^\text{ c}_1,\bm{o}^\text{ c}_2,\cdots,\bm{o}^\text{ c}_T\right]$ is the output of last-layer Transformer, $\bm{e}_u$ is the embedding for the K-means unit $u$, and $\tau$ scales the logit, set to 0.1. Similar to the speaker loss in speaker information modeling, the content loss $\mathcal{L}_\text{ c}$ is only applied over the masked frames.

\subsubsection{Loss Function of Pre-training}
As mentioned in section~\ref{pre_sec}, our progressive residual extraction method works effectively only when all modules are trained jointly. Furthermore, the speaker extractor needs to be co-supervised by a pre-trained speaker-teacher model. Our \our{} model is pre-trained using the following multi-task loss function:
\begin{equation}
\mathcal{L} = \lambda_\text{ f} \cdot \mathcal{L}_\text{ f} + \lambda_\text{ s} \cdot \mathcal{L}_\text{ s} + \lambda_\text{ c} \cdot \mathcal{L}_\text{ c},
\end{equation}
where $\mathcal{L}_\text{ f}$ denotes the mean square error of $\bm{X}_f$. $\mathcal{L}_\text{ s}$ and $\mathcal{L}_\text{ c}$ represent the losses associated with speaker and content modeling, respectively, as described in equation~(\ref{spk_loss}) and (\ref{hubert_loss}). The hyper-parameters $\lambda_\text{ f}$, $\lambda_\text{ s}$, and $\lambda_\text{ c}$ for the three loss functions are set to 10.0, 1.0, and 1.0, respectively.

\subsection{Fine-tuning}
As introduced in Section \ref{pf}, after pre-training, we utilize a weighted-sum mechanism for downstream fine-tuning, as depicted in Fig.~\ref{proposed}. All outputs of the hidden layers are weighted-sum with learnable weights as input to the downstream model. Due to the insertion of two specific task extractors, \our{} utilizes the representations extracted by the pitch extractor, speaker extractor, and the outputs of Transformer layers (excluding the layers with inserted two extractors) for weighted-sum, as shown in Fig.~\ref{proposed}. This approach enables us to use different weights to obtain representations suitable for various downstream tasks. Consequently, with a lightweight downstream model, we can achieve excellent performance on downstream tasks with a small amount of supervised data.

\section{Experiments and Results}
\label{experiment_sec}
\subsection{Tasks and Datasets}
We pre-trained our model on LibriSpeech \cite{librispeech}, WenetSpeech \cite{wenetspeech}, and Multi-lingual Speech (MLS) \cite{pratap2020mls}. We conducted various fine-tuning experiments, including speech recognition (ASR), speaker identification (SID), speech enhancement (SE), emotion recognition (ER), and voice conversion (VC) to evaluate the model's performance on content, speaker, intonation, and acoustic learning. These fine-tuning experiments utilized data from various datasets: LibriSpeech \cite{librispeech}, VoxCeleb1 \cite{voxceleb1}, Interactive Emotional Dyadic Motion Capture (IEMOCAP) \cite{busso2008iemocap}, Voicebank-DEMAND \cite{sedata}, and the dataset of the Voice Conversion Challenge (VCC2020) \cite{vcdata}. All audio samples were sampled at 16 kHz.
	
\textbf{Pre-training}: We pre-trained two versions of \our{} and HuBERT: Base and Large. For the Base model, we used 960 hours of LibriSpeech data for pre-training to ensure comparability with other open-source Base SSL models. For the Large model, we used a total of 84,500 hours of bilingual data in English and Chinese for pre-training. This bilingual dataset included 44,500 hours of MLS English data, 10,000 hours of Wenetspeech Chinese data, and 30,000 hours of Chinese speech data collected from the Internet. All pre-training data were used without labels.

\textbf{Speech recognition fine-tuning}: The \textit{train-clean-100} and \textit{dev-clean} subsets of LibriSpeech were employed as the training and development datasets for ASR, respectively. The performance of the models was evaluated on the \textit{test-clean}, and \textit{test-other} subsets of LibriSpeech.

\textbf{Speaker identification fine-tuning}: We fine-tuned and evaluated the models on the VoxCeleb1 dataset for the SID task. VoxCeleb1 contains over 100,000 utterances from 1,251 celebrities, extracted from videos.

\textbf{Speech enhancement fine-tuning}: We used the Voicebank-DEMAND dataset for SE. This dataset includes data from 28 speakers with various signal-to-noise ratio (SNR) levels: 15, 10, 5, and 0 dB. The test set consists of data from two additional speakers with 17.5, 12.5, 7.5, and 2.5 dB SNRs.

\textbf{Speech emotion recognition fine-tuning}: We fine-tuned models on section 2 to 5 subsets of the IEMOCAP dataset for ER and evaluated their performance on the section 1 subset. The IEMOCAP dataset comprises approximately 12 hours of recordings encompassing various emotional expressions. Notably, IEMOCAP emphasizes the natural expression of emotions in conversations, where the emotional content of speech is closely tied to the spoken context.

\textbf{Voice conversion fine-tuning}: Following the evaluation in SUPERB-SG \cite{s3prl_vc, superbsg}, we conducted the any-to-one voice conversion task of VCC2020, where TEF1 was chosen as the target speaker. The speaker model was directly trained on the target speaker training set.


\subsection{Experimental Setup}

\subsubsection{Configuration of Models}

To validate the effectiveness of our proposed \our{}, we conducted comprehensive comparisons with some excellent self-supervised pre-training models as follows:

\textbf{wav2vec 2.0} \cite{wav2vec2}: wav2vec 2.0 is a self-supervised learning model that utilizes a quantization-based contrastive learning strategy to pretrain the encoder, distinguishing between positive samples (audio segments from the same utterance) and negative samples (segments from different utterances).

\textbf{WavLM} \cite{wavlm}: WavLM simultaneously learns the BERT-like masked unit prediction and denoising during pre-training. It has shown SOTA performance on various downstream tasks.

\textbf{HuBERT} \cite{hubert}: HuBERT is a self-supervised speech representation learning approach that employs an offline clustering step to provide aligned target pseudo labels for a BERT-like prediction loss.

\textbf{\our{}}: Our proposed progressive residual extraction based pre-training strategy is illustrated in Fig.~\ref{proposed}. Two extractors are inserted into the encoder-style SSL backbone. The pitch extractor consists of three 256-channel convolutional layers with a kernel size of 5, a single-layer GRU with 256 cells, and a fully connected (FC) layer with hidden feature-dimension cells. The speaker extractor comprises a fully connected layer with 256 cells, an FAS layer with hidden feature-dimension cells, and another fully connected layer with hidden feature-dimension cells. The hidden feature dimension is 768 in the Base model and 1024 in the Large model.

We compared the Base and Large versions of the four models, keeping the parameter configurations of the main structures consistent. For the Base version model, the convolutional module consists of seven layers, each with 512 channels, with strides of $\left\{5,2,2,2,2,2,2\right\}$ and kernels of $\left\{10,3,3,3,3,2,2\right\}$. The Transformer contains 12 layers with 768 dimensions, 3072 inner dimensions, and 12 attention heads. In contrast, for the Large version model, the convolutional module maintains the same configuration as the Base version, but its Transformer contains 24 layers with 1024 dimensions, 4096 inner dimensions, and 16 attention heads. 

Our pre-training codebase is built on MindSpore, resulting in a slight loss\footnote{\url{https://github.com/wangtianrui/ProgRE/blob/master/supplementary_results/README.md\#migration-errors}} in accuracy when migrating the model to various downstream fine-tuning tasks implemented in the PyTorch framework. To distinguish our baseline from HuBERT$_{\text{pt}}$, which is pre-trained in PyTorch \cite{ott2019fairseq}, we refer to our baseline HuBERT, implemented under the MindSpore framework \cite{mindspore}, as HuBERT$_{\text{ms}}$ or baseline.

\subsubsection{Pre-training Setup}
The pseudo labels, speaker teacher, and detailed settings for pre-training are introduced as follows:

\textbf{Unsupervised unit discovery}: In our model's pre-training process, we conduct two iterations, with the primary distinction being the origin of pseudo-labels for the main branch. During the first iteration, we extract 13-dimensional MFCCs along with their first-order and second-order differential features. Subsequently, we train a 100-class K-means model using the resulting 39-dimensional features from 10\% (1\% for Large) of the speech data. Finally, we assign the corresponding cluster center as the pseudo-label for each frame of speech. In the second iteration, we utilize the output of the middle layer of the model pre-trained in the first iteration as features (the 9th layer for the Base version and the 18th layer for the Large version). These features are then used to train a 500-class K-means model, and the corresponding cluster center is assigned as the pseudo-label for each frame of speech. For clustering, we utilize the MiniBatchKMeans implemented in the scikit-learn \cite{kramer2016scikit} with a mini-batch strategy. We set the mini-batch size to be 10,000 frames. Additionally, we employ k-means++ \cite{arthur2007k} with 20 random starts for better initialization.

\textbf{Self-supervised speaker teacher model}: We employed the open-source toolkit Wespeaker \cite{wang2023wespeaker} to pre-train the EMA-DINO \cite{dino} without labels as the speaker teacher model for our \our{}. Specifically, for the Base version of \our{}, we trained the EMA-DINO model with 512 intermediate dimensions on 960 hours of LibriSpeech. Similarly, for the Large version of \our{}, we trained the teacher model with 1024 intermediate dimensions on the 44,500 hour MLS English dataset. These teacher models will output a 192-dimensional utterance-level speaker embedding to serve as supervision for the speaker extractor of our \our{}.

\textbf{Training detail}: For the Base version, with two iterations, \our{} Base was pre-trained for 400K steps per iteration on 32 Ascend910 GPUs, with a batch size of 60-second samples per GPU. For the Large version, \our{} Large was pre-trained for 350K steps in the first iteration and 1400K steps in the second iteration, using 96 Ascend910 GPUs with a batch size of 25-second samples per GPU. The Adam optimizer was used with a warm-up learning rate, ramping up from 0 to 5e-4 for the first 8\% of steps and then decaying to 0.

\subsubsection{Fine-tuning Setup}
The downstream models and detailed settings for fine-tuning are introduced as follows:

\textbf{Downstream models}: For ASR, we used a 2-layer BiLSTM with 1024 cells, optimized by the character-unit CTC loss. For SID, we applied an utterance-level mean-pooling followed by a 1251-class FC layer, optimized by cross-entropy loss. For SE, we used a 3-layer BiLSTM with 256 cells followed by a sigmoid activation for mask-based filtering, trained via the L1 loss function. For ER, an utterance-level mean-pooling followed by convolutional attention with a kernel size of 5 was optimized by cross-entropy loss. For VC, we used the Taco2-AR model\footnote{\url{https://github.com/s3prl/s3prl/blob/main/s3prl/downstream/a2o-vc-vcc2020/config.yaml}}. Taco2-AR extracts a speaker vector via an encoder consisting of 3 convolution layers and a 1024-cell BiLSTM and then generates the log-mel spectrograms using 2-layer 256-cell LSTM models in an auto-regressive manner.

\textbf{Fine-tuning detail}: To effectively evaluate the capabilities learned by the self-supervised pre-trained models, we froze the parameters of the pre-trained model during fine-tuning and only fine-tuned the weights of the downstream model and the weights of the weighted-sum mechanism. All downstream fine-tuning was performed using the Adam optimizer. Due to the varying data scales of different downstream tasks, the number of training steps, learning rate, and batch size used for fine-tuning each downstream task differed. The detailed configuration is shown in the TABLE~\ref{config_downtask}.
\begin{table}[htp]
    \centering
    \caption{Fine-tuning configuration of downstream tasks.}
    \begin{tabular}{ccccc}
        \toprule
        Task  & update step & learning rate  & batch size  \\
        \midrule
        ASR & 40k & 1e-4 & 32  \\
        SID & 100k & 1e-3 & 64   \\
        SE &  40k & 1e-3 & 16  \\
        ER &  50k & 1e-4 & 16   \\
        VC &  10k & 1e-4 & 6  \\
        \bottomrule
    \end{tabular}
    \label{config_downtask}
\end{table}

\subsubsection{Metrics}

Word error rate (WER) was used to evaluate performance in the speech recognition task. Accuracy (Acc) was employed for speaker identification (SID) and speech emotion recognition (SER). Perceptual Evaluation of Speech Quality (PESQ) \cite{pesq} and scale-invariant signal-to-distortion ratio (SI-SDR) \cite{sisdr} were used to measure the quality of enhanced speech with a clean reference. Higher PESQ scores indicate better auditory quality of the enhanced speech, and higher SI-SDR values indicate greater similarity between the clean and enhanced signal distributions. For the voice conversion (VC) task, we used Mel-Cepstral Distortion (MCD) \cite{mcd}, Pearson correlation coefficient of pitch (F0C) \cite{cohen2009pearson}, WER, and speaker accept rate (SPK) for evaluation. WER was measured using a pre-trained ASR model\footnote{\url{https://dl.fbaipublicfiles.com/fairseq/wav2vec/wav2vec_vox_960h_pl.pt}}, and SPK was defined as the pass rate at which the speaker verification model\footnote{\url{https://github.com/resemble-ai/Resemblyzer}} considered the converted speech to be consistent with the target speaker.

\subsection{Ablation Study}
We first verify the effectiveness of each improvement in our model. We conduct ablation experiments involving the residual extraction, speaker extractor, and pitch extractor. In these experiments, we focus on the model's performance on two tasks: speech recognition and speaker identification. These tasks evaluate the model's ability to understand speech content and non-linguistic information, respectively \cite{information}.

\subsubsection{Importance of Residual Extraction}
Residual extraction is the core of our method. Based on the Base version model, we compared the performance of residual extraction with that of multi-task extraction, where multi-task extraction replaces the subtraction (denoted by $\circleddash$) in residual extraction with addition (denoted by $\oplus$). Since the insertion layer of the speaker extractor also affects the performance of the model, in this ablation experiment, we inserted the speaker extractor at the position where it performs best in the Base version setting, which is after the 4th layer of the Transformer, results are shown in the TABLE~\ref{abs_res}.

\begin{table}[htp]
    \setlength\tabcolsep{4.5pt}
    \setlength{\extrarowheight}{2.5pt}
    \centering
    \caption{Comparison of residual extraction and multi-task extraction. \textbf{BLOD} indicates the best result. 
    }
    \begin{tabular}{c|c|cc|cc}
        \cline{1-6}
        \multirow{2}*{Index}  & \multirow{2}*{Method}  & \multicolumn{2}{c|}{ASR (WER) $\downarrow$} & \multicolumn{2}{c}{SID (Acc) $\uparrow$}  \\
        \cline{3-6}
        &&   test-clean  & test-other & dev & test   \\
        \cline{1-6}
        0 &baseline & 6.85 & 16.77 & 81.01 & 79.94  \\
        \cline{1-6}
        1 &$\oplus$ pitch $\oplus$ speaker & 8.06  & 19.11 & 88.75 & 87.58   \\
        2 &$\circleddash$ pitch $\oplus$ speaker & 7.87  & 18.55 & 89.69 & 89.14  \\
        3 &$\oplus$ pitch $\circleddash$ speaker & 6.71  & 16.36 & 88.77 & 87.51  \\
        4 &$\circleddash$ pitch $\circleddash$ speaker & \textbf{6.52}  & \textbf{15.20} & \textbf{90.95} & \textbf{90.61}   \\
        \cline{1-6}
    \end{tabular}
    \label{abs_res}
\end{table}

Although multi-task extraction ($\oplus$ pitch $\oplus$ speaker) significantly enhances the pre-trained model's ability to extract speaker information, it degrades the model's performance on speech recognition. This occurs because multi-task extraction strengthens the model's capacity to capture pitch variation and speaker information, but the enhanced content-irrelevant information interferes with the deeper Transformer's capacity to extract content information. 
When we remove the enhanced pitch variation information from the main branch ($\circleddash$ pitch $\oplus$ speaker), the performance of speech recognition improves, and the improvement of speaker identification becomes more significant. This is because pitch variation information is less related to speaker and content information, and removing redundant information can enhance performance on irrelevant tasks jointly.
The 3-index method ($\oplus$ pitch $\circleddash$ speaker) performs comparably to multi-task extraction (1-index) on speaker identification tasks, but its performance on speech recognition is improved compared to the 2-index method. This indicates that the speaker information extracted by the speaker extractor has a more significant interference with content extraction than pitch variation information.
The final results ($\circleddash$ pitch $\circleddash$ speaker) demonstrate that progressive residual extraction can jointly enhance the model's performance on both speech recognition and speaker identification tasks. Progressively refining the content information in the main branch improves the model's performance across various tasks.

\subsubsection{Importance of Inserting the Speaker Extractor}
Unlike the pitch extractor, which directly takes its input from the waveform, the speaker extractor is inserted after the middle Transformer layer. Therefore, we conducted an ablation experiment on the insertion layer. In this ablation experiment, we used the Base version model, did not include the pitch extractor, and used residual extraction (subtraction, $\circleddash$) as the insertion method. We inserted the speaker extractor before the Transformer (0th layer) or after the $\{2,4,6,8,10,12\}$-th layer. The results are shown in TABLE~\ref{spk_ablation}.

\begin{table}[htp]
    \centering
    \setlength{\extrarowheight}{2.5pt}
    \caption{Comparison of the insertion layer of speaker extractor. \textbf{BLOD} indicates the best result.}
    \begin{tabular}{c|cc|cc}
        \cline{1-5}
        \multirow{2}*{Layer}  & \multicolumn{2}{c|}{ASR (WER) $\downarrow$} & \multicolumn{2}{c}{SID (Acc) $\uparrow$}  \\
        \cline{2-5}
        &   test-clean  & test-other & dev & test   \\
        \cline{1-5}
        - & 6.85 & 16.77 & 81.01 & 79.94  \\
        \cline{1-5}
        0 & 7.01  & 17.02 & 73.16 &  72.17  \\
        2 & 7.46  & 17.61 & 77.59 & 75.37   \\
        4 & \textbf{6.67}  & \textbf{16.04} & \textbf{88.89} & \textbf{87.64}  \\
        6 & 7.48  & 18.30 & 86.88 &  85.37 \\
        8 & 8.01  & 19.41 & 74.36 & 72.55  \\
        10 & 8.17  & 21.27 & 73.26 &  69.54 \\
        12 & 7.74 & 18.73 & 84.61 & 84.25  \\
        \cline{1-5}
    \end{tabular}
    \vspace{-0.3cm}
    \label{spk_ablation}
\end{table}

\begin{table*}[!htp]
    \setlength\tabcolsep{4.5pt}
    \setlength{\extrarowheight}{2.5pt}
    \centering
    \caption{Comparison of pre-training methods fine-tuned on different downstream tasks.}
    \begin{tabular}{lc|c|c|cc|cc|cc|c|cccc}
        \cline{1-15}
        \multicolumn{2}{c|}{\multirow{2}*{Method}}  & \multirow{2}*{\makecell[c]{Param.\\(M)}} & \multirow{2}*{Codebase}  & \multicolumn{2}{c|}{ASR (WER) $\downarrow$} & \multicolumn{2}{c|}{SID (Acc) $\uparrow$} & \multicolumn{2}{c|}{SE} & \multirow{2}*{\makecell[c]{ER\\(Acc) $\uparrow$}} & \multicolumn{4}{c}{VC} \\
        \cline{5-10} \cline{12-15}
        & & & & test-clean  & test-other & dev & test & PESQ $\uparrow$ &  SI-SDR $\uparrow$& & MCD $\downarrow$ & F0C $\uparrow$ & WER $\downarrow$  & SPK $\uparrow$   \\
        \cline{1-15}
        \multirow{5}*{\rotatebox{90}{Base}}&wav2vec2.0& 94.70 & Fairseq & 6.75 & 16.28 & 79.08 & 78.22 & 2.94 & 9.35 & 62.21 & 7.86 & 0.35 & 11.2 & 92.0 \\
         &HuBERT$_{\text{pt}}$& 94.70 & Fairseq & 6.72 & 16.11 & 81.42 & 80.17 & 2.99 & 9.32 & 62.44 & 7.89 & 0.34 & 10.2 & 94.0 \\
         &WavLM& 94.70 & Fairseq & \textbf{6.50} & 15.27 & 84.58 & 83.59 & 3.00 & 9.08 & 63.78 & \textbf{7.74} & 0.39 & 9.9 & 94.0  \\
         &HuBERT$_{\text{ms}}$& 94.70 & Mindspore & 6.85 & 16.77 & 81.01 & 79.94 & 2.96 & 9.24 & 62.05 & 7.77 & 0.35 & 10.4 & 94.0 \\
         &\textbf{\our{}}& 97.04 & Mindspore & 6.52 & \textbf{15.20} & \textbf{90.95} & \textbf{90.61}  & \textbf{3.04} & \textbf{9.80} & \textbf{63.96} & 7.75 & \textbf{0.40} & \textbf{8.8} & \textbf{97.0}  \\
         \cline{1-15}
         \multirow{5}*{\rotatebox{90}{Large}}&wav2vec2.0& 317.38 & Fairseq & 3.87 & 8.80 & 91.25 & 90.54 & 3.01 & 9.28 & 67.82 & 8.20 & 0.16 & 16.7 & 87.0 \\
         &HuBERT$_{\text{pt}}$& 317.38 & Fairseq & 3.96 & 8.82 & 92.41 & 92.29 & 3.03 & 9.41 & 69.49 & 7.77 & 0.36 & 11.3 & 90.0 \\
         &WavLM& 317.38 & Fairseq & \textbf{3.79} & \textbf{8.26} & 96.47 & 96.08 & \textbf{3.11} & 9.44 & 70.69 & 7.86 & 0.35 & 11.2 & 92.0\\
         &HuBERT$_{\text{ms}}$ & 317.38 & Mindspore & 4.07 & 9.41 & 90.49 & 90.38 & 3.00 & 9.26 & 67.13 & 7.84  & 0.34 & 11.5 & 90.0 \\
         &\textbf{\our{}}& 319.72 & Mindspore & 3.86 & 8.64 & \textbf{97.67} & \textbf{97.61} & 3.09 & \textbf{9.45} & \textbf{70.73} & \textbf{7.74} & \textbf{0.39} & \textbf{9.5} & \textbf{94.0} \\
        \cline{1-15}
    \end{tabular}
    \vspace{-0.5cm}
    \label{main_table}
\end{table*}

The results indicate that the insertion of the speaker extractor at different layers significantly impacts the model's performance. Compared to the baseline, inserting the speaker extractor before the Transformer (0th layer) or after the $\{2,8,10\}$-th layers leads to a degradation in the model's performance on speaker identification. 
This outcome can be attributed to the task characteristics of different Transformer layers, as the main branch employs the self-supervised strategy of HuBERT. 
Previous works \cite{wavlm, layer_ana} analyzed the task characteristics of different layers and found that the fourth, fifth, sixth, and eleventh layers play significant roles in speaker tasks, with the fourth layer being the most influential. 
Hence, strengthening and removing that does not align with the task characteristics of the main branch will instead degrade the model's performance on that task.
Regarding the 0th layer, since our lightweight speaker extractor lacks temporal modeling, it cannot effectively extract speaker information from the output after local-processing convolution. Furthermore, inserting the speaker extractor after the $\{0,2,6,8,10,12\}$-th layers leads to a decline in the main branch's training efficacy, as observed in the degraded performance on the speech recognition task. In summary, the effectiveness of inserting residual extraction relies on the task characteristics of different Transformer layers obtained from the main branch training strategy.

\subsubsection{Importance of Inserting the Pitch Extractor}
We explored the role of the pitch extractor in the Base version of our model, and the results are presented in TABLE~\ref{pitch_ablation}. The findings indicate that strengthening and then removing pitch variation can improve the model's performance in both speech recognition and speaker identification tasks. This improvement can be attributed to the fact that pitch variation primarily contains intonation information, with minimal speaker and content information \cite{vqmivc}. Consequently, removing pitch variation information from the main branch facilitates the subsequent extraction of speaker and content information.

\begin{table}[h]
    \setlength\tabcolsep{2pt}
    \centering
    \setlength{\extrarowheight}{2.5pt}
    \caption{Evaluation of inserting the pitch extractor.}
    \begin{tabular}{l|c|cc|cc}
        \cline{1-6}
        \multirow{2}*{Method} & \multirow{2}*{\makecell[c]{Param.\\(M)}} & \multicolumn{2}{c|}{ASR (WER) $\downarrow$} & \multicolumn{2}{c}{SID (Acc) $\uparrow$}  \\
        \cline{3-6}
        & &  test-clean  & test-other & dev & test   \\
        \cline{1-6}
         baseline & 94.70 & 6.85 & 16.77 & 81.01 & 79.94  \\
        $\circleddash$ pitch  & 95.47 & 6.74  & 16.38 & 81.66 & 80.03  \\
        $\circleddash$ speaker & 96.27 & 6.67  & 16.04 & 88.89 & 87.64  \\
        $\circleddash$ speaker $\circleddash$ pitch & 97.04& \textbf{6.52}  & \textbf{15.20} & \textbf{90.95} & \textbf{90.61}  \\
        \cline{1-6}
    \end{tabular}
    \vspace{-0.3cm}
    \label{pitch_ablation}
\end{table}

\subsection{Comparing SSL Models on Various Downstream Tasks}
In order to verify the performance of our proposed method on various downstream tasks, we conducted a comparison with existing open-source pre-trained models on speech recognition, speaker identification, speech enhancement, emotion recognition, and voice conversion tasks. The results are shown in TABLE~\ref{main_table}. Comparing the results of HuBERT$_{\text{ms}}$ and HuBERT$_{\text{pt}}$, it shows that the pre-trained HuBERT based on the MindSpore loses accuracy when migrating to PyTorch, resulting in slight performance degradation on each downstream task. Despite this disadvantage, our proposed method still demonstrates SOTA performance on most tasks. Note that HuBERT$_{\text{ms}}$ is used as our baseline instead of HuBERT$_{\text{pt}}$. In addition, for the Large version of \our{}, we inserted the speaker extractor after the 6th layer Transformer.

\subsubsection{Speech Recognition}
We compared the models' ability to content understanding via fine-tuning models on speech recognition. 

In the Base version models, compared to HuBERT$_\text{ms}$, our \our{} achieves a relative WER reduction of 8.04\%, our \our{} even outperforms WavLM implemented by Fairseq on test-other, indicating that residual extraction of the pitch variation and speaker information can effectively facilitate the learning of irrelevant content information. 

In the large version models, in addition to our proposed \our{}, we also pre-trained HuBERT$_{\text{ms}}$ based on the MindSpore framework on our 84,500 hours of bilingual English-Chinese data for comparison. 
The results indicate that the 84,500 hours of bilingual data did not bring a significant improvement in performance compared with LibriLight's 60,000 hours. 
The amount of English data used in pre-training is 15,500 hours less than that of HuBERT$_{\text{pt}}$. Coupled with the limitations of the MindSpore framework, HuBERT$_{\text{ms}}$'s ASR ability in English is worse than that of HuBERT$_{\text{pt}}$. Although additional experiments\footnote{\url{https://github.com/wangtianrui/ProgRE/blob/master/supplementary_results/README.md\#chinese-asr}} showed that HuBERT$_{\text{ms}}$ performs better than HuBERT$_{\text{pt}}$ in Chinese ASR, it is unfair to compare this to Fairseq models that have not seen Chinese data, so those results are not shown in this paper. Despite these challenges, the proposed model still shows excellent performance, second only to WavLM pre-trained on 94,000 hours of English-only pertaining data, which proves that the progressive residual extraction strategy is also applicable to large-scale pre-training tasks.

\subsubsection{Speaker Identification}
We assessed the model's capacity to extract speaker information through speaker identification tasks. The results demonstrate that our proposed method achieves SOTA performance in both the Base and Large version groups, showcasing substantial enhancements compared to HuBERT$_{\text{ms}}$. This improvement in speaker information extraction can be attributed to the residual insertion of the speaker extractor at an optimal position, which ensures the effectiveness of main branch training while significantly boosting the insertion layer's ability to extract speaker information.

\begin{figure*}[t]
\centering
\vspace{-0.2cm}
\includegraphics[width=18.5cm]{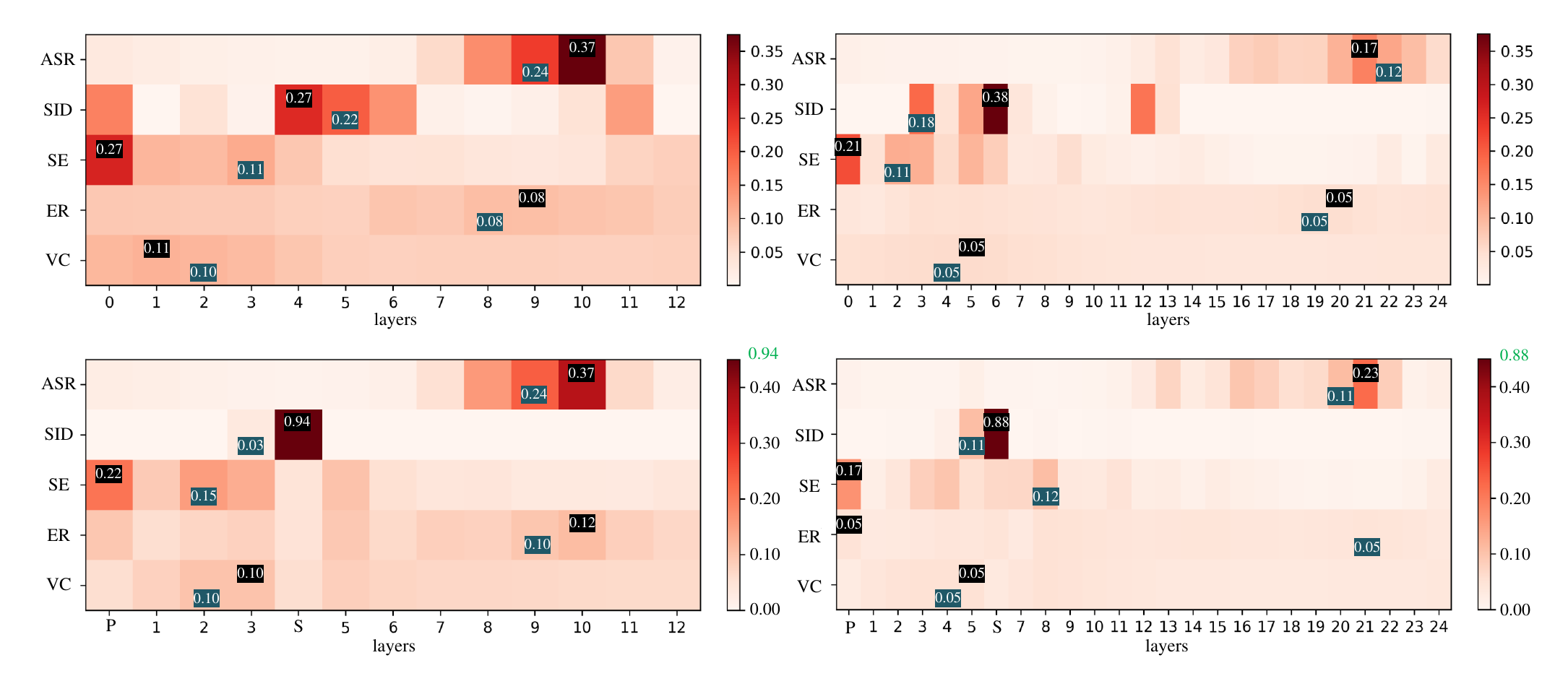}
\vspace{-1.1cm}
\caption{Layer-wise weight visualization in the weighted-sum mechanism of HuBERT and \our{}. The first row weights come from HuBERT, and the second row comes from \our{}. We show weights fine-tuned on ASR and SID tasks for both Base and Large version models (left column is Base version, right column is Large version).} 
\label{layer_ana}
\vspace{-0.5cm}
\end{figure*}

\subsubsection{Speech Enhancement}
Speech enhancement requires extracting clean and detailed acoustic information from noisy input, so the ability of the pre-trained model to extract various speech information is comprehensively evaluated. The results show that our \our{} achieves the best performance among the Base version models. This is because WavLM Base does not introduce denoising during pre-training, thus \our{}, with its superior content information extraction, speaker information extraction, and pitch information extraction abilities, achieves the best speech enhancement performance among the Base models. In the Large version models, the overall PESQ metric is improved compared to the Base, but the SI-SDR has decreased. We speculate that this is because the features extracted by the Large model are more abstract with refined semantic information, making the enhanced speech more intelligible to human ears but causing distortion in numerical acoustic details. WavLM Large introduced denoising during pre-training, and combined with its excellent performance in the speech recognition task, it achieves the highest PESQ score. The PESQ score of our proposed method is slightly lower than that of WavLM Large. However, by strengthening the extraction of pitch and speaker information closer to the speech signal, our method \our{} Large achieves a slightly higher SI-SDR score compared to WavLM Large.

\subsubsection{Speech Emotion Recognition}
Since emotive expression and content are interrelated in the IEMOCAP dataset \cite{busso2008iemocap}, the performance of speech emotion recognition reflects both the model's ability to content understanding and its ability to extract paralinguistic information. The results show that the proposed model achieves the best performance under both Base and Large configurations. This improvement can be attributed to the residual extraction of pitch variation information, which allows the model to capture more intonation and tonal details while also extracting refined and accurate content information.

\subsubsection{Voice Conversion}
\label{sec_vc}
We conducted an any-to-one voice conversion experiment \cite{s3prl_vc}. The model needs to remove the intonation and speaker information from the input speech and then generate the speech of the target speaker based on the remaining content information, thus serving as a test of the model's capability to disentangle speech information. In addition to objective evaluation, we also present audio samples on the demo page\footnote{\url{https://wangtianrui.github.io/progre_vc}}. From the results, we can see that \our{} achieves a significantly higher speaker verification pass rate (SPK) and F0 correlation (F0C) compared to other reference models, indicating that the content representations extracted by \our{} contain fewer speaker and intonation information (an in-depth analysis can be found in Section~\ref{layer_ana_sec}). This proves the effective information removal of residual extraction in \our{}. Moreover, the superior performance on the WER metric suggests that the \our{}-based VC model can retain more complete content information. The SOTA overall performance is due to \our{} extracting pitch variation and speaker information residually at optimal layers, making the intonation, speaker, and content information extracted by our model more independent, enhancing the disentanglement of speech information.

Unexpectedly, the performances of the Large models are generally slightly worse than those of the Base models. We speculate that this is because the downstream model is too small to fit the high-dimensional features extracted by the Large models, making it difficult for the model to converge. Nevertheless, our proposed method still shows the best performance among the Large models, indicating that the content information extracted by our proposed model is more refined and easier for the downstream model to learn.

\subsection{Layer-wise Weight in Weighted-sum Mechanism}
\label{layer_ana_sec}
In order to explore the task characteristics of features at different layers, we visualized the weights in the weighted-sum mechanism, as shown in Fig.~\ref{layer_ana}. Since the numerical distribution of our proposed method is extreme, we cropped the value by 0.45 when drawing the weights of \our{}, as shown in green text. To more intuitively show the difference in values, we marked the top-2 weights. 
As can be seen from Fig.~\ref{layer_ana}, consistent with the findings in other papers, the HuBERT model exhibits different task characteristics at different layers. 
For content understanding tasks such as speech identification, the weights of the features extracted by the deep Transformer layers are high. Conversely, for the extraction of non-linguistic information such as speaker recognition, the weights of the features extracted by the shallow layers play major roles. 
Compared with HuBERT, the weights of our method on the speaker identification task are concentrated in the layer where the speaker extractor is inserted, and the weights of the other layers are close to zero. This demonstrates that residual extraction can effectively remove information from the main branch, allowing subsequent layers to focus on speaker-irrelevant tasks. 
Additionally, the weights of the proposed method on the speech recognition task are similar to the weight distribution of the original HuBERT, further proving that adding residual extraction of corresponding task information at the appropriate layer does not change the task characteristics distribution of the main branch layer for other irrelevant tasks.
For speech enhancement, the overall weight distribution of \our{} is similar to that of HuBERT, with weights gradually decreasing from shallow to deep layers. For emotion recognition, pitch variation representation and content-related layers play key roles. This is because pitch variation contains emotion-related intonation information. For voice conversion, the shallow-layer weights of HuBERT and \our{} are slightly larger than deep-layer's weights. However, the weights of the pitch representation and speaker feature layers in \our{} are low. Combined with the results of Section~\ref{sec_vc}, our method can reduce the intonation and speaker information in the weighted-sum representation by decreasing the weights of the pitch and speaker extractor layers. This also implies that the intonation and speaker information contained in the representations of other layers is less than that of HuBERT.


\section{Discussion}
\label{discussion}

In this study, we explored the effectiveness of residual extraction in speech self-supervised pre-training, highlighting that the task characteristics of different layers facilitated by the pre-training strategy are crucial for achieving joint improvement across various downstream tasks. 
Previous research \cite{giuliani2006improved} has shown that actively normalizing speaker information in speech recognition models can effectively enhance performance on ASR. Consistent with these findings, our experimental results demonstrate that actively removing content-irrelevant speech information from the main branch, such as speaker information or pitch variation, can improve the model's ability to extract content information. 
This validates that residual extraction is essential for a single SSL model to adapt to various downstream tasks with diverse demands for different types of speech information. Moreover, as illustrated in TABLE~\ref{spk_ablation} and Fig.~\ref{layer_ana}, the strengthening of pre-training model tasks should align with the layer-specific task characteristics facilitated by the main branch pre-training strategy. 
Deviating from these task characteristics can detrimentally affect the main branch's training efficacy, resulting in degraded performance. 
To verify the practicality of our proposed method, we expanded the dataset to 84,500 hours of English-Chinese bilingual data. The experimental results demonstrate that the proposed method is adaptable to large-scale pre-training tasks, achieving joint performance improvements across various tasks. Notably, \our{} exhibits state-of-the-art performance in speaker information extraction and speech information disentanglement capabilities. 
This confirms our hypothesis: to make the pre-training model more universal, it is essential to enhance specific capabilities while minimizing the interference of these strengthened features with other irrelevant tasks. Our findings provide a significant reference for the development of universal pre-training models.

This paper also points out some interesting issues that need further exploration. First, after additional experimental verification, we found that our model trained under the MindSpore framework incurs a numerical relative error of 0.25\% when being migrated to the PyTorch framework, slightly affecting the model's performance during fine-tuning. To address this, we have open-sourced our code to encourage other researchers to try it under the PyTorch framework. Second, for the Large version model, we used 44,500 hours of English data and 40,000 hours of low-quality Chinese data. From the comparative experiment of HuBERT$_{\text{pt}}$ and HuBERT$_{\text{ms}}$, we observed that although our total dataset is larger than that of HuBERT$_{\text{pt}}$, which was trained with 60,000 hours of English-only data, the performance in downstream fine-tuning tasks for English is declined. We speculate that speech quality and language differences significantly impact the performance of pre-trained models. Third, for a fair comparison, the speaker-teacher model in our \our{} is limited to a self-supervised model on the pre-training data. Exploring the possibility of directly using a powerful supervised model is an attractive direction. Finally, this paper focuses on the speech pre-training model, whether the concept of progressive residual extraction is applicable to audio pre-training in general is another interesting issue.

\section{Conclusion}
\label{conclusion}

Improving performance on various downstream tasks jointly is a challenge that has garnered significant attention in speech self-supervised pre-training. In this paper, we highlight that different downstream tasks require different types of speech information. To make an SSL model more universal, it is crucial to mitigate the mutual interference of irrelevant speech information extraction during pre-training. Inspired by pitch-speaker-content decoupling in voice conversion and speaker information normalization in speech recognition, we propose a progressive residual extraction based pre-training method. By leveraging the task characteristics of different layers in HuBERT's self-supervised strategy, we enhance specific layers' abilities to extract pitch variation and speaker information. Subsequently, we remove this enhanced information from the main branch using a residual extraction approach. This removal reduces the subsequent learning burden on content extraction for the main branch, ultimately achieving joint improvements across various downstream tasks.


\bibliographystyle{IEEEtran}
\bibliography{references}

\end{document}